\documentclass[%
 reprint,
 longbibliography,
superscriptaddress,
 amsmath,amssymb,
 aps,
]{revtex4-1}
\usepackage{graphicx}% Include figure files
\usepackage{dcolumn}% Align table columns on decimal point
\usepackage{bm}% bold math
\newcommand{\ud}{\mathrm{d}}
\usepackage{subfigure}  % use for side-by-side figures
\begin{document}

\preprint{APS/123-QED}

\title{Pairing of Fermions with Unequal Effective Charges in an Artificial Magnetic Field}% Force line breaks with \\

\author{F. Nur \"{U}nal}
\email{fatmanur@bilkent.edu.tr}%
\affiliation{Laboratory of Atomic and Solid State Physics, Cornell University, Ithaca, New York 14853, USA}
\affiliation{Department of Physics, Bilkent University, 06800 Ankara, Turkey}
\author{M. \"{O}. Oktel}
\affiliation{Department of Physics, Bilkent University, 06800 Ankara, Turkey}


%\affiliation{
%\affiliation{
%\affiliation{%


\date{\today}% It is always \today, today,

\begin{abstract}
Artificial magnetic fields (AMFs) created for ultra cold systems depend sensitively on the internal structure of the atoms. In a mixture, each component experiences a different AMF depending on its internal state. This enables the study of Bardeen-Cooper-Schrieffer pairing of fermions with unequal effective charges. In this Letter, we investigate the superconducting (SC) transition of a system formed by such pairs as a function of field strength. We consider a homogeneous two-component Fermi gas of unequal effective charges but equal densities with attractive interactions. We find that the phase diagram is altered drastically compared to the usual balanced charge case. First, for some AMFs there is no SC transition and isolated SC phases are formed, reflecting the discrete Landau level (LL) structure. SC phases become reentrant both in AMF and temperature. For extremely high fields where both components are confined to their lowest LLs, the effect of the charge imbalance is suppressed. Charge asymmetry reduces the critical temperature even in the low-field semiclassical regime. We discuss a pair breaking mechanism due to the unequal Lorentz forces acting on the components of the Cooper pairs to identify the underlying physics.
\begin{description}
\item[PACS numbers]
\end{description}
\end{abstract}

\pacs{Valid PACS appear here}% PACS, the Physics and Astronomy
\maketitle

%\tableofcontents

Cold atom experiments have realized novel many-particle systems, challenging some of the most fundamental models of condensed matter theory. In particular, Bardeen-Cooper-Schrieffer (BCS) theory of fermion pairing, which has successfully explained superconducting (SC) and superfluid (SF) behavior in a large number of systems, had to be extended to cover new regimes. Pairing due to resonant interactions has been explored both theoretically and experimentally, uncovering the BEC-BCS crossover in detail. Density imbalance between components forming the Cooper pairs and resulting unconventional SC states were first considered for condensed matter systems, but, experimental observation of polarized SFs \cite{KetterleImbalance,HuletImbalance} with cold atoms required significant improvements upon prior approaches. Similarly, Cooper pairs made up of fermions with unequal masses have been explored theoretically \cite{Wilczek,Caldas}.

The constituents of cold atom experiments are neutral atoms. The dominant interaction between these atoms are through s-wave scattering which can be tuned via Feshbach resonances between the atoms. While the absence of Coulomb interactions facilitated the realization of some fundamental condensed matter models, the neutrality of the particles prevented the observation of the effects of an external magnetic field on these systems. Initial efforts in this direction used rotation to mimic the magnetic field which brings further constraints on the confining potential of the ultracold system \cite{CooperRotation}.

Over the last five years, a significant development, namely the creation of Raman laser-assisted artificial magnetic fields (AMFs) for neutral atoms \cite{spielman}, has extended the capabilities of cold atom experiments. These AMFs are realized by coupling the internal states of the atoms to light to imprint a Berry phase on the motion. While a number of different schemes have been used to manufacture these synthetic Hamiltonians, all of them sensitively depend on the internal excitation structure \cite{DalibardArtificialMag.}. Hence, for a mixture of two different atom species or even a mixture composed of atoms in different hyperfine states, the effective magnetic field acting on each component can be different. For example, the \textit{g}-factors for $^{87}$Rb $5S_{1/2}$ $F=1$ and $^{85}$Rb $5S_{1/2}$ $F=2$ have a $3/2$ ratio. If the scheme in Ref.\cite{spielman} is applied to a mixture of these atoms, position dependent detunings, consequently the AMFs, would reflect this ratio. Although the Zeeman shifts due to the real magnetic field are utilized to create the AMF, this artificial field only couples to the spatial motion of the atoms and does not cause an artificial Zeeman effect \cite{DalibardArtificialMag.}.

In this Letter, we explore the consequences of an AMF that couples unequally to the fermions forming a Cooper pair. Essentially, we consider the pairing of fermions with different cyclotron frequencies, which we regard as unequal effective charges coupling to the same AMF. We discuss the conditions for pairing and the response of the paired state to the external AMF. We show that this system displays reentrant SC in temperature, \textit{i.e.} a normal sample at zero temperature can become SC as temperature is increased. Oscillatory dependence of $T_C$ on the AMF, which is a direct consequence of the Landau Level (LL) structure of single-particle excitations, is observed. However, for some AMFs, SC state is not preferred even at zero temperature. We calculate the phase diagram of the system for various representative cyclotron frequency ratios and present physical mechanisms to elucidate the fundamental changes in the SC transition.

We consider a mixture of two fermion species of equal mass and equal density. The system is assumed to be spatially homogeneous, as in most cold atom experiments the effects of the confining potential can be taken into account through the local density approximation. An AMF of arbitrary strength is acting on the system by coupling only to the orbital motion of the fermions but causing no Zeeman shift. The coupling of the AMF to each component is different, defining the effective charges $q_1$ and $q_2$. Corresponding cyclotron frequencies $\omega_1=q_1B/m$ and $\omega_2=q_2B/m$ define the respective LL separations. We introduce the relative frequency $\omega_r=\omega_2/\omega_1$ and the effective frequency $\omega=\sqrt{\omega_1\omega_2}$. Within the Landau gauge $\vec{A}=(0,Bx,0)$, the non-interacting Hamiltonian can be written as
\begin{equation}
\mathcal{H}_0(\omega_1,\omega_2)=\sum_{\nu}\varepsilon_{1\nu}f_{1\nu}^{\dag}f_{1\nu}^{\phantom{\dag}}+\varepsilon_{2\nu}f_{2\nu}^{\dagger}f_{2\nu}^{\phantom{\dagger}},
\end{equation}
where the index $\nu=(n,k_y,k_z)$ incorporates the LL index $n$, momentum along the \textit{z}-direction $k_z$, and the momentum $k_y$ which also labels the LL degeneracy. The associated kinetic energy is $\varepsilon_{i\nu}=\epsilon_{i\nu}-\mu_i=\hbar^2k_z^2/2m+\hbar\omega_i(n+\frac{1}{2})-\mu_i$, with $i=1,2$. The chemical potentials $\mu_i$ are not equal but are chosen to fix the density of both species to be the same at each AMF value, $N_1=N_2=N/2=Bq_i/2\pi h\sum_n\int\ud k_z F(\varepsilon_{i\nu})$ with the Fermi-Dirac distribution function $F(\varepsilon_{i\nu})$. The particle densities are then scaled with the effective magnetic length, $\ell=\sqrt{\hbar/m\omega}$, i.e. $n_1=N\pi^2\ell^3$. We numerically solve the number equations for chemical potentials at each AMF. Thus, for a fixed charge ratio $\omega_r$ and total real-space density $N$, changing the AMF strength alters only the effective density $n_1$.

This single particle spectrum is unique as the LLs of up and down spins do not match in energy. Since their zero point energy and the separation between the LLs are different, two LLs can have equal energy only if the charge ratio $\omega_r$ is a rational number. When the chemical potentials are adjusted to equate the densities, low energy single-particle excitation spectra for up and down spins are asymmetric. This mismatch has drastic consequences on pairing when the interactions are introduced.

The two species interact resonantly through s-wave scattering which we model using the two-channel Hamiltonian following Refs.\cite{MC,HZ} studying the balanced magnetic field case, $\omega_r=1$. We write the interacting Hamiltonian,
\begin{multline}
\mathcal{H}(\omega_1,\omega_2)=\sum_{\nu}\varepsilon_{1\nu}f_{1\nu}^{\dag}f_{1\nu}^{\phantom{\dag}}+\varepsilon_{2\nu}f_{2\nu}^{\dagger}f_{2\nu}^{\phantom{\dagger}} +\varepsilon_Bb^{\dag}b^{\phantom{\dag}} \\ +\alpha\sum_{\nu\nu\prime}Q_{\nu\nu\prime}^*b^{\dag}f_{1\nu}^{\phantom{\dag}}f_{2\nu\prime}^{\phantom{\dag}}+H.c.\qquad
\end{multline}
The open channel fermions interact to form the closed channel boson with energy $\varepsilon_B=\gamma+(\hbar\omega_1+\hbar\omega_2)/2-\mu_1-\mu_2+C$, where $\gamma$ is the unrenormalized detuning between the closed channel boson and the open channel fermions. C is the counterterm which is set to compensate the divergence of the boson self energy due to the infinite number of LLs of the closed channel fermions,
\begin{multline}
C=-\frac{\alpha^2m}{4\pi\hbar^2}\sum_{n,n^{\prime}}\frac{(n+n^{\prime})!}{n!n^{\prime}!}\frac{\ell_1^{2n+1}\ell_2^{2n^{\prime}}}{(\ell_1^2+\ell_2^2)^{n+n^{\prime}+1}}\\
\times\frac{1}{\sqrt{n+\frac{1}{2}+\omega_r(n^{\prime}+\frac{1}{2})}}.
\end{multline}
The closed channel boson eigenstates have the same form as the fermion single particle states, however the effective charge of the boson is $q_B=q_1+q_2$, and its mass $2m$. Consequently the boson magnetic length is $\ell_B=\ell_1\ell_2/\sqrt{\ell_1^2+\ell_2^2}=\sqrt{\hbar/m(\omega_1+\omega_2)}$. Two fermions with up and down spin in different LLs can interact to form a boson through the coupling constant $\alpha$ and the overlap integral $Q_{\nu\nu\prime}$ which has the form
\begin{multline}
Q_{\nu\nu^{\prime}}=\delta(k_z+k_{z^{\prime}})\frac{(-1)^n\ell_1^n\ell_2^{n^{\prime}}}{\sqrt{\sqrt{\pi}2^{n+n^{\prime}}n!n^{\prime}!(\ell_1^2+\ell_2^2)^{n+n^{\prime}+1/2}}}\\ \times e^{-\frac{k_y^2}{2}(\ell_1^2+\ell_2^2)} H_{n+n^{\prime}}\Big(k_y\sqrt{\ell_1^2+\ell_2^2}\Big),\quad
\end{multline}
where $H_n(.)$ is the \textit{n}-th Hermite polynomial. The dominant pairing mechanism is through the closed channel boson with the lowest possible energy. Hence, the wave function of the boson is in general a superposition over all the states in the bosonic lowest LL (LLL) with zero $k_z$. The distribution of the boson wave function over the degenerate LLL states with different $k_y$ does not affect the physical properties after the bosons are integrated out \cite{HZ,MC}. Thus, we calculate the overlap integral only for the LLL state with $k_y=0$. The renormalization parameters $\gamma$ and $\alpha$ are chosen in such a way that the system produces low energy scattering properties for $(\omega_1,\omega_2)\rightarrow0$. Hence, they are related to the physical parameters $a_s$\textit{-scattering length} and $r_0$\textit{-effective range} as follows,
\begin{eqnarray}
-\frac{\gamma}{\alpha^2}=\frac{m}{a_s\hbar^24\pi}, \qquad\quad \frac{1}{\alpha^2}=-\frac{r_0m^2}{8\pi\hbar^4}.
\end{eqnarray}

In order to analyze the pairing of unequal charges, we examine this Hamiltonian around the SC transition. Following mean-field theory, we introduce $\alpha<b>=\Delta$ which is the average amplitude of the bosonic wave function defining the order parameter. Near the transition, $\Delta$ is small and we expand the free energy,
\begin{multline}
\mathcal{F}=\mathcal{F}_0+\frac{\varepsilon_B}{\alpha^2}|\Delta|^2\\
-\frac{1}{2}\sum_{\substack{n,n^{\prime}\\k_y,k_z}}|\Delta|^2 |Q_{nn^{\prime}}|^2 \frac{\tanh\frac{\varepsilon_{1n}}{2k_BT}+\tanh\frac{\varepsilon_{2n^{\prime}}}{2k_BT}}{\varepsilon_{1n}+\varepsilon_{2n^{\prime}}}. \label{Free en.}
\end{multline}
Pairing is favorable if the coefficient of the second order term is negative. The critical temperature for pairing is obtained by setting this coefficient to zero
\begin{multline}
-\frac{1}{a_s}=\frac{\hbar^2}{m}\sum_{n,n^{\prime}}^{\infty}\frac{(n+n^{\prime})!}{n!n^{\prime}!}\frac{\ell_1^{2n}\ell_2^{2n^{\prime}}}{(\ell_1^2+\ell_2^2)^{n+n^{\prime}+1}} \\\times\int_{-\infty}^{\infty}\frac{\ud k_z}{2\pi}\Bigg\{\frac{\tanh\frac{\varepsilon_{1n}}{2k_BT}+\tanh\frac{\varepsilon_{2n^{\prime}}}{2k_BT}}{\varepsilon_{1n}+\varepsilon_{2n^{\prime}}}-\frac{2}{\epsilon_{1n}+\epsilon_{2n^{\prime}}} \Bigg\}.\label{Gap Eq}
\end{multline}
The chemical potentials $\mu_i$ are chosen so that the real-space densities of the two components are the same. The right-hand side of Eq.\ref{Gap Eq}, the pairing susceptibility, determines the behavior of $T_C$ and can be used to understand the underlying physical picture for its evolution.

We solve Eq.\ref{Gap Eq} numerically by calculating the pairing susceptibility for a given value of T and $n_1$. The phase diagram of the system is then obtained by comparing this value with $-1/a_s$. For the numerical solution we scale all energies by the effective magnetic energy $\hbar\omega$. Similarly the dimensionless scattering length is $\tilde{a}_s=a_s(N\pi^2)^{1/3}$. Our equations are symmetric for $\omega_r\rightarrow\frac{1}{\omega_r}$ which is equivalent to switching the indices of the components. We checked this symmetry numerically and concentrate on $0<\omega_r\leq1$ in the following.

\begin{figure}
\includegraphics[width=0.47\textwidth]{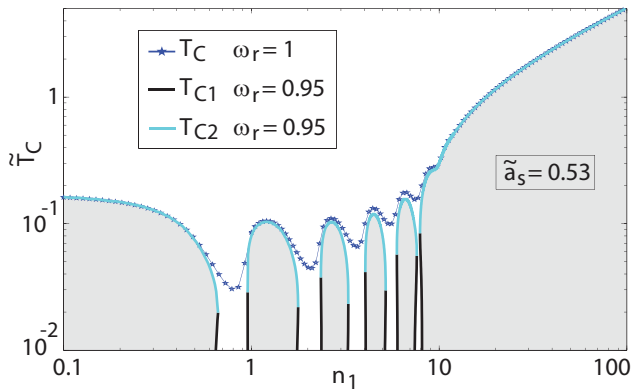}
\caption{(Color online) Phase diagram of the system as a function of dimensionless temperature $\tilde{T}=k_B T/\hbar\omega$ and effective density $n_1=N\pi^2\ell^3$ where $N$ is the real-space density and $\ell=\sqrt{\hbar/m\omega}$. Notice that increasing AMF, $B=m\omega/\sqrt{q_1q_2}$, corresponds to lower $n_1$ values. Two frequency ratios $\omega_r=1$ and $\omega_r=0.95$ are displayed for the same interactions strength $\tilde{a}_s=a_s(N\pi^2)^{1/3}=0.53$. For $\omega_r=1$, there is SC transition at any field, the oscillations in $T_C$ (stars) originates from the LL structure. For $\omega_r=0.95$, these oscillations evolve into bubble SC regions (shaded areas). The system is not SC even at zero temperature between the bubbles and the transition becomes weakly reentrant in temperature. At the low (many LLs) and high field (only LLL) regimes, $T_C$ is not affected significantly by a small charge imbalance.  } \label{log phase Fig.}
\end{figure}

\begin{figure}
\centering
\subfigure{\includegraphics[width=0.23\textwidth]{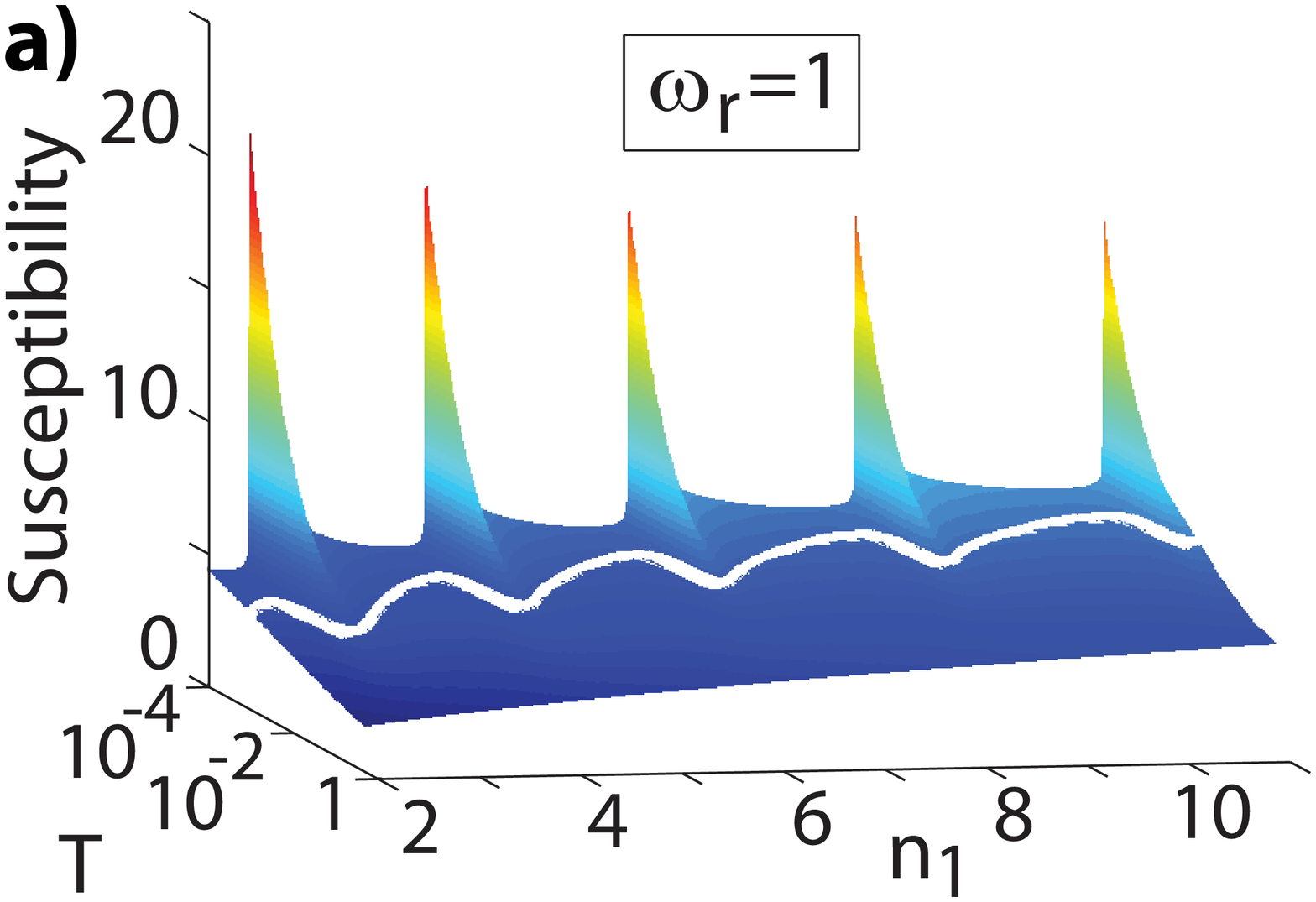} \label{s1 fig}}
\subfigure{\includegraphics[width=0.22\textwidth]{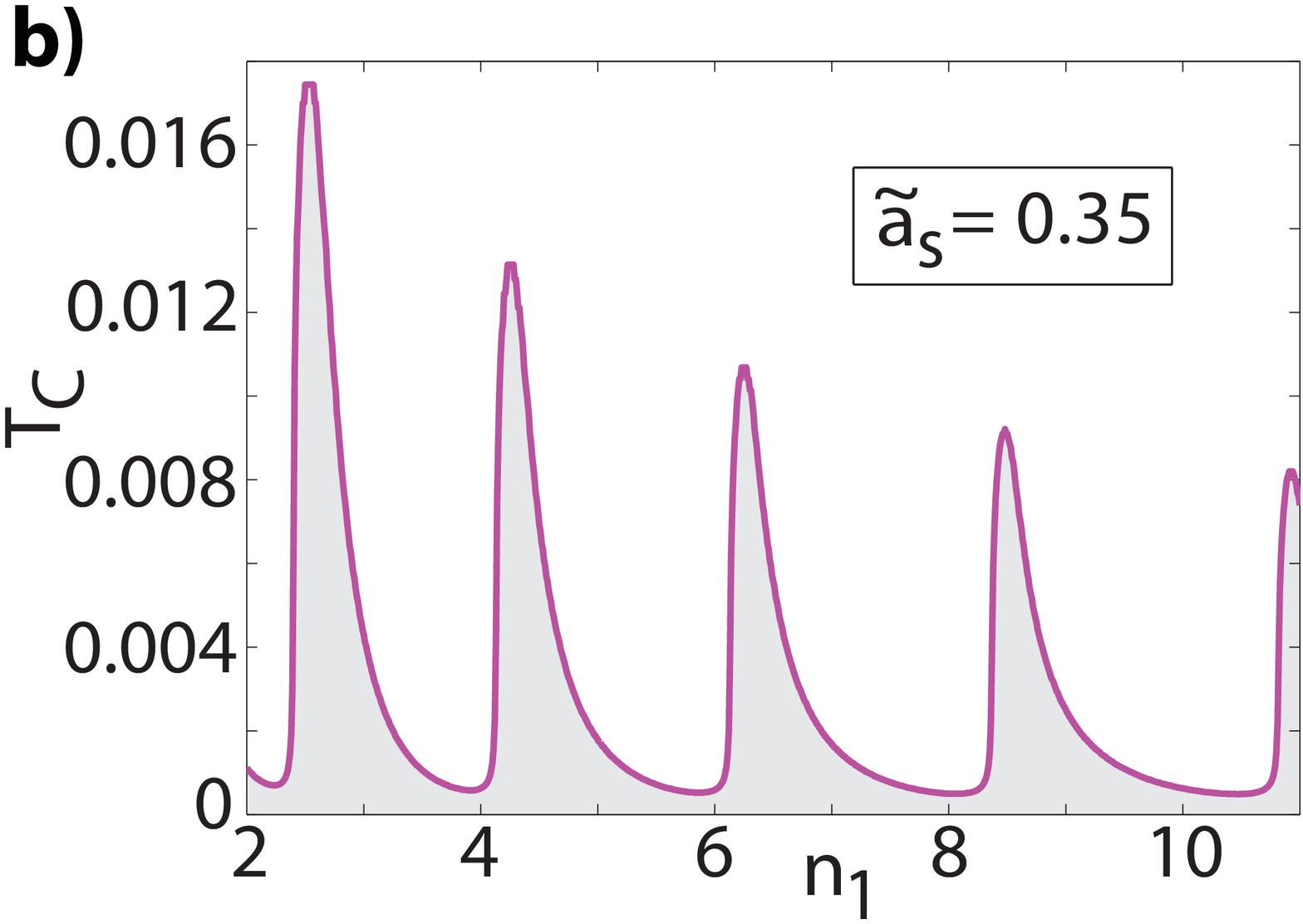} \label{p1  fig}}
\subfigure{\includegraphics[width=0.23\textwidth]{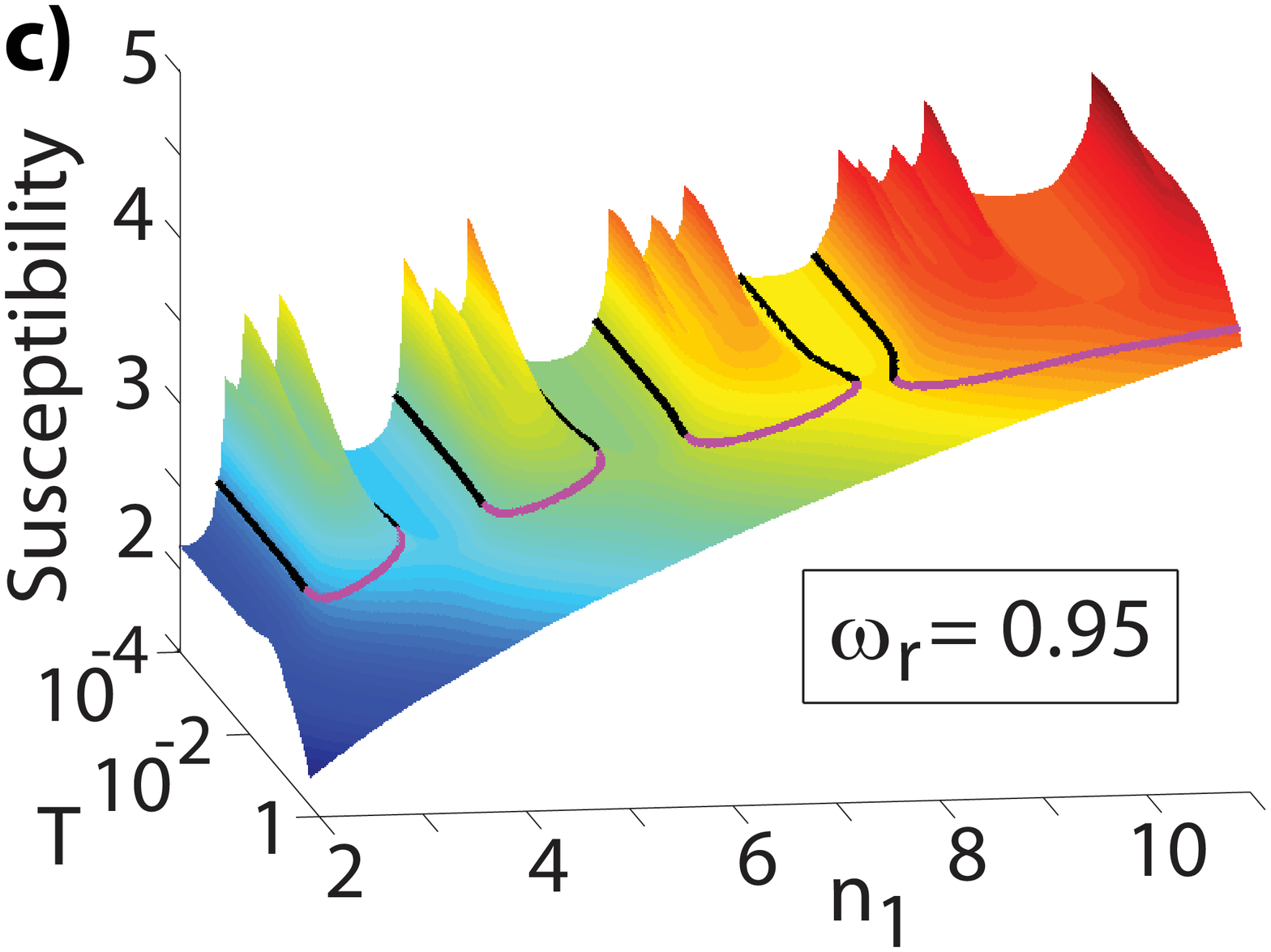} \label{s95  fig}}
\subfigure{\includegraphics[width=0.22\textwidth]{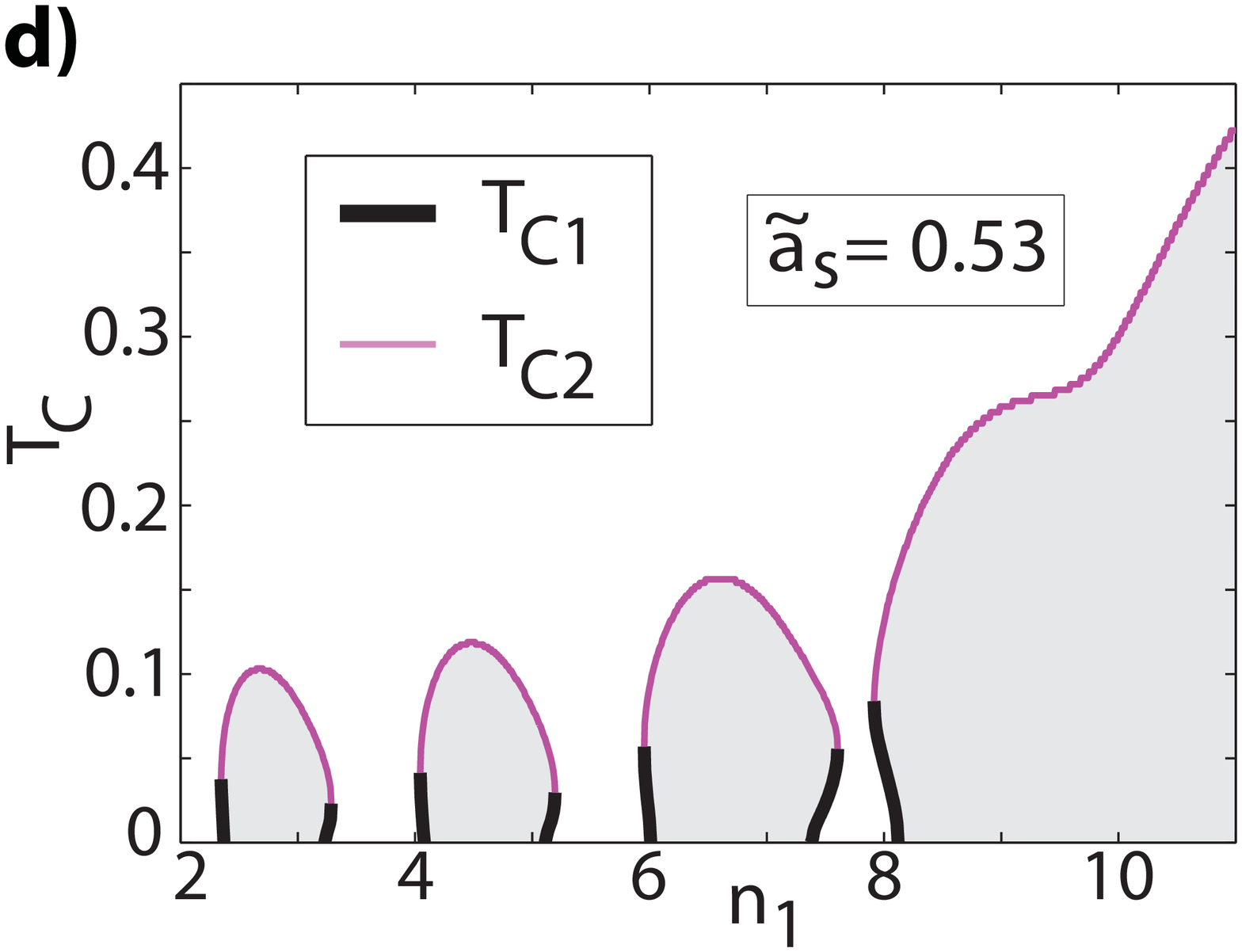} \label{p95  fig}}
\subfigure{\includegraphics[width=0.23\textwidth]{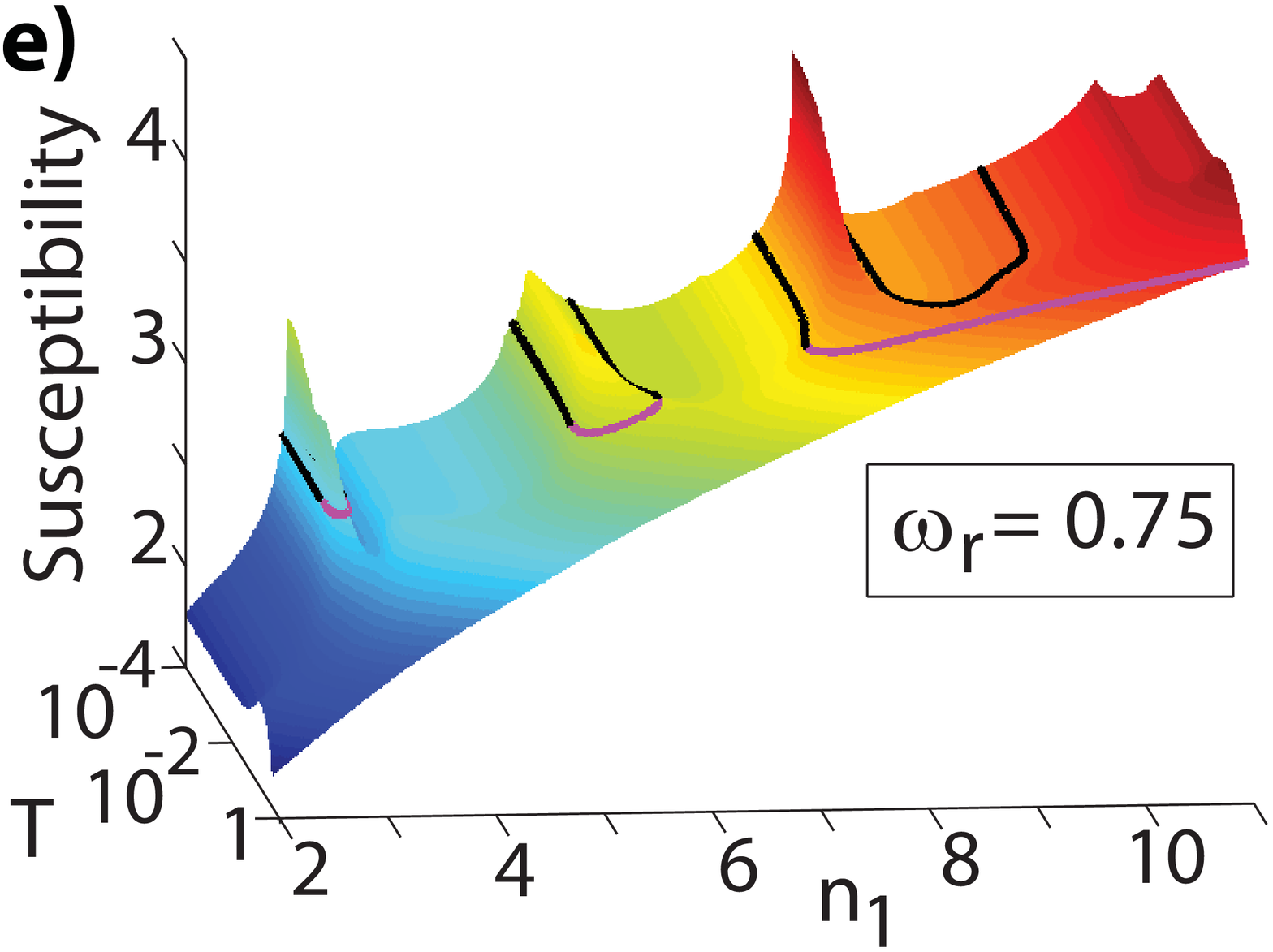} \label{s75  fig}}
\subfigure{\includegraphics[width=0.22\textwidth]{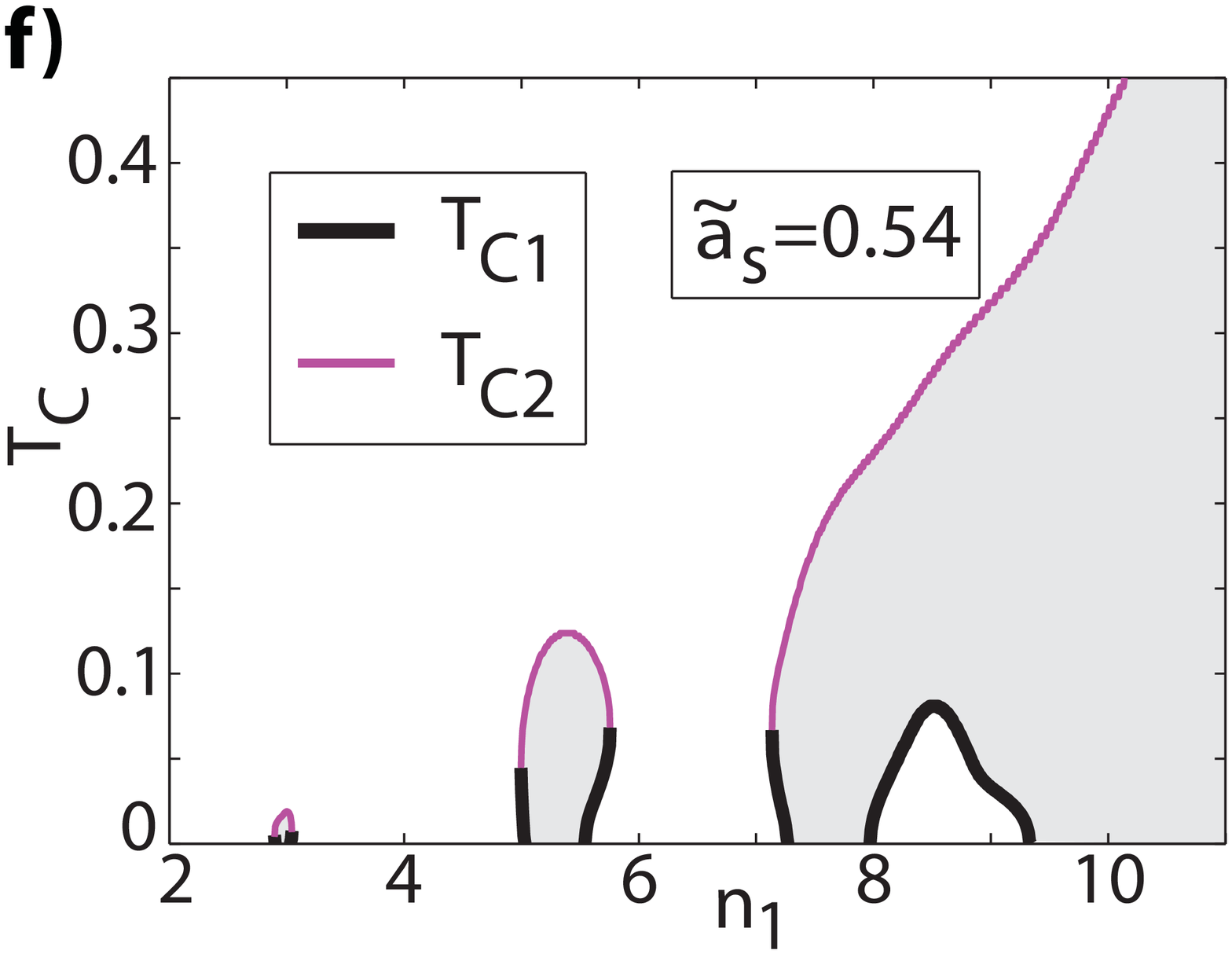} \label{p75 fig}}
\caption{ (Color online) Pairing susceptibility (left panel) and respective phase diagrams (right panel) for three frequency ratios $\omega_r=1,0.95,0.75$. The system is made dimensionless with effective magnetic energy $\hbar\omega$ as in Fig.\ref{log phase Fig.} and decreasing $n_1$ corresponds to increasing AMF at fixed real-space density $N$. The phase diagrams are in linear scale in order to cover $T_C=0$ and plotted for intermediate field strengths where charge imbalance effects are most prominent. a) Pairing susceptibility of equal charges diverges at low temperature, guaranteeing SC for any field value. Divergence is more pronounced at LL thresholds. The corresponding phase diagram (b) is obtained by Eq.\ref{Gap Eq}. The phase boundary is also highlighted on the surface. c) Even a slight asymmetry between the charges, $\omega_r=0.95$, lifts the low temperature divergences and the oscillations in (b) turn into bubble SC phases (d). Each LL susceptibility peak is split into smaller peaks, thus, the bubble phases branch into smaller bubbles for weaker interactions (not displayed). e,f) The mismatch between LL spectra is greater for smaller $\omega_r$, resulting in prominent reentrance with temperature. The maximum reentrance temperature is controlled by $\hbar|\omega_1-\omega_2|$. } \label{wr=1,95,75 figs}
\end{figure}

In Fig.\ref{log phase Fig.}, we present the phase diagram of $\omega_r=1$ which is in agreement with Ref.\cite{MC}. For the balanced charge case, there is always a critical temperature below which the SC state is preferred within the mean-field approximation. The critical temperature is non-monotonic with the applied field. It first decreases and becomes exponentially small as smaller number of LLs are involved in the pairing, but, then increases when only the LLL contributes. This high-field SC has been studied for both solid state \cite{Tesanovic,TesanovicPRL} and cold atom systems \cite{MC}. The oscillatory behavior of $T_{C}$ with the applied field is a direct result of the underlying LL spectrum. When the LLs of the two components coincide in energy, the pairing susceptibility diverges as $1/\sqrt{T}$ at the peaks and as $\ln(T)$ at the rest in Fig.\ref{s1 fig}, always guaranteeing a SC state.

We also display the phase diagram for $\omega_r=0.95$ in Fig.\ref{log phase Fig.}. First of all, unlike the balanced case, there are AMF values for which a SC state is never favored. Oscillatory behavior in pairing due to the LL structure causes stable islands of SC in the phase diagram isolated from the zero field SC phase. Although the low field SC is not affected from unequal frequencies as can be expected, the high-field SC proves to be surprisingly resilient too. The destruction of the low temperature SC in the presence of a small misalignment between LLs has been predicted \cite{GuntherGruenberg}, however, this misalignment also leads to a third effect, reentrant SC. For some AMFs, at low temperatures the sample is in the normal state while at higher temperatures it becomes SC. Consequently, even for a slight asymmetry between the charges of the components, the phase diagram undergoes fundamental changes. These changes are most pronounced in the regime where only a few LLs are populated for both components and we display representative phase diagrams in Fig.\ref{wr=1,95,75 figs}. In the following, we discuss the physical reason for each of these three features.

Fig.\ref{s95  fig} displays the pairing susceptibility for $\omega_r=0.95$ by focusing on intermediate field strength. The most striking feature of this phase diagram is the emergence of isolated islands of SC Fig.\ref{p95  fig} which come about because Eq.\ref{Gap Eq} does not have a solution at some AMF strengths. The absence of solutions even for a minute amount of charge imbalance is best understood by considering the single-particle spectra. The one-particle density of states (DOS) for each component has sharp peaks at each LL threshold. For the balanced case, the DOS and the chemical potentials of the components are always equal. If the temperature is low enough, only the DOS near the chemical potential is relevant. At low temperatures, the pairing susceptibility diverges as $T^{-1/2}$ at each LL threshold and the peaks in $T_C$ follow from the one-particle DOS. When $\omega_r\neq1$, LLs of different components do not have the same energy or total degeneracy. In general, chemical potentials of both components must be chosen differently to give equal real-space densities. If the mismatch between the chemical potentials and the LLs is large, the energy cost of exciting particles may not be redeemed by attractive interactions even at zero temperature.

The pockets of SC phases roughly correspond to population of a new LL. Whenever a chemical potential crosses a LL threshold, there is a new set of states at the Fermi surface which become suddenly available to contribute to the pairing. The most favorable case for pairing is when both chemical potentials simultaneously cross a new LL. For small charge imbalance, these threshold crossings happen within a small difference in AMF which essentially turns the $T_C$ oscillations of the balanced case into bubble SC phases in Fig.\ref{p95  fig}. For a general $\omega_r$ ratio, the picture is much more complicated. For chemical potentials to give equal densities and cross LL thresholds simultaneously, $\omega_r$ must be close to a simple fraction. This complicated behavior is evident in the pairing susceptibilities displayed in Fig.\ref{wr=1,95,75 figs} where simple peaks of balanced LLs are split into smaller structures. For very strong attractive interactions, these bubble phases are not resolved as $T_C$ becomes comparable to the LL separation. A similar effect was predicted for the balanced case \cite{MC}.

As the AMF is increased further, we observe that $T_C$ increases and reaches the same value with the equal charge limit. This surprising revival happens only when both components populate their LLLs. Although the degeneracies of the two LLLs are not equal, they both increase with increasing AMF. The chemical potentials of both components then lie very close to the corresponding LLL thresholds. Hence, the excitation cost for pairing decreases at such high fields. We can estimate the AMF for which the effect of charge imbalance vanishes by requiring all the particles with smaller effective charge to reside in their LLL. This estimate is in good agreement with our numerical results.

Another fundamental change brought about by the charge imbalance is SC that is reentrant with temperature. While this effect is not clear for small imbalance as in Fig.\ref{p95  fig}, we found that it is a common feature of the phase diagrams for general $\omega_r$. A more prominent reentrant SC phase can be observed for $\omega_r=0.75$ as in Fig.\ref{p75 fig}. For some field strengths, the system prefers normal phase at zero temperature and becomes SC only above $T_{C1}$. SC phase subsequently disappears after a higher temperature $T_{C2}$. Similar reentrant behavior was predicted for graphene bilayers \cite{Zareyan} and asymmetric nuclear matter \cite{LombardorReentrantSC}. In our system it is easy to understand the physical basis for this reentrance. Increasing temperature generally prefers a disordered state, however, it also excites a significant amount of particles to higher LLs. Because of the antisymmetric and oscillatory nature of the DOS in the charge imbalanced system, states which are close to not only the chemical potential but also to LL thresholds are most favorable for pairing. Thus, if the pairing contribution from thermally excited particles overcomes the entropy cost, increasing the temperature can drive the SC transition. With this scenario we expect the maximum lower critical temperature $T_{C1}$ to be of the order of LL mismatch between the components which agrees with our numerical results. In contrast to other reentrant SC phases where a competing order precludes SC at low temperatures \cite{fulde}, the current system has reentrance solely due to non-trivial nature of the single-particle DOS.

Although we concentrate on the interplay between discrete LL structure and charge asymmetry, it is worth mentioning that there is a profound effect even in the semiclassical regime where many LLs are filled for both components. The effect of an external magnetic field on a Cooper pair is modeled by only considering the phase acquired by the center of mass motion. This semiclassical approximation due to orbital dephasing describes the upper critical field $H_{C2}$ in type-II SC successfully. However, if the charges of the fermions forming the pairs are different, the magnetic field couples the center of mass motion with the relative coordinate. As pairing is controlled by the relative coordinate, especially for tightly bound pairs, it is possible for the Lorentz force due to the center of mass motion to break the pairs. Classically, if the center of mass of a bound pair with charges $q_1,q_2$ is moving with velocity $v$ in a perpendicular magnetic field $B$, Lorentz force difference between the two particles is $F\simeq|q_1-q_2|Bv$. This real space picture can be utilized to give a rough estimate for the strength of this pair breaking mechanism by comparing the work done by this force over the size of the pair to the SC gap. This effect becomes dominant especially for large charge ratios and we estimate the upper critical field due to this pair breaking mechanism as $H_{C3}\approx\frac{\sqrt{\omega_r}}{1-\omega_r}H_{C2}$ which is in agreement with our numerical results. While competing orders such as charge density waves, which were not taken into account in our approach, may complicate the physics in the high magnetic field limit, the decease of $T_C$ with charge imbalance is observed even when many LLs are filled and the mean-field approximation is most reliable.

In summary, the cold atom experiments with AMFs can create mixtures where each component has a different effective charge. The pairing between fermions of unequal effective charges presents a unique extension of BCS theory which is fundamental in diverse areas of physics. In this Letter, we find that even for a slight asymmetry between the charges, the phase diagram changes drastically with the emergence of reentrant SC both in temperature and AMF. The oscillatory behavior of $T_C$ with AMF for the balanced case modifies into isolated SC phases. For extremely high AMFs where both components are in their LLLs, the transition temperature is independent of the charge ratio. Finally, we argue that $T_C$ is reduced due to pair breaking facilitated by unequal Lorentz forces on the charges forming the pairs.

F. N. \"{U}. is supported by T\"{u}rkiye Bilimsel ve Teknolojik Ara\d{s}t{\i}rma Kurumu (T\"{U}B\.{I}TAK). This work is supported by T\"{U}B\.{I}TAK Grant No. 112T974.

\bibliography{PairingPRLResubmit3}% Produces the bibliography via BibTeX.

\end{document}